\documentstyle[11pt,subeqn]{article}
\begin{document}
\hbadness=10000

\pagestyle{plain}
\baselineskip 21.pt

%\input def.tex
%\input bold.tex

%\ni\today \hfill{\bf }
\begin{center}
{\LARGE\bf MEASUREMENT IN QUANTUM PHYSICS}

\bigskip

Michael Danos~\footnote{Visiting Scholar.}\\
Enrico Fermi Institute,\\
University of Chicago, Chicago, Illinois 60637, USA\\
and\\
Tien D. Kieu~\footnote{Present address: CSRIO, Australia.}\\
School of Physics,\\
University of Melbourne, Parkville, Victoria 3052, Australia

\end{center}

\bigskip

\begin{abstract}
\noindent
The conceptual problems in quantum mechanics -- related to the collapse of the wave function, the particle-wave duality, the meaning of measurement -- arise from the need to ascribe particle character to the wave function.  As will be shown, all these problems dissolve when working instead with quantum fields, which have both wave and particle character.  Otherwise the predictions of quantum physics, including Bell's inequalities, coincide with those of the conventional treatments. The transfer of the results of the quantum measurement to the classical realm is also discussed. 

\end{abstract}

\section{Generalities}
A vast literature exists on the interpretation of
quantum mechanics in general, and on the meaning of measurement in
quantum mechanics in particular. The discussion still takes place
today; from small, semi-popular papers~\cite{1} to highly technical
large-scale programmatic treatments~\cite{2,2'}.  To
quote Bell, Ref.~\cite{3}, who in discussing some of his articles
in that book writes in the preface: ``these [articles] show my
conviction that, despite numerous solutions of the [measurement]
problem  ..., a problem of principle remains.''  (See also Feynman,
Ref.~\cite{F}.) The most fundamental of the problems interfering with the
understanding of quantum mechanics is indeed the problem of
measurement, the ``genesis of information,'' and all the effects
surrounding what has been termed ``the  Copenhagen collapse of the
wave function,'' which is not described by the Schr\"odinger
equation \cite{PP}.  This process not only lies outside of the framework of
quantum mechanics, but, being instantaneous, also violates
relativistic causality.   As we will see, all of this is closely
related to the \hbox{so-called}  ``particle-wave duality'' and
is the source of Bell's above mentioned ``problem of principle.'' Several
different proposals to overcome Bell's above ``problem of
principle'' have been made.  They all in some way break
the framework of quantum theory.  We shall not discuss these proposals,
but refer the reader to the above mentioned
literature citations~\cite{1,2,2',3,PP}.

Our point is that all problems associated with the subject known as ``the
problem of measurement in quantum mechanics'' {\it can be resolved without
abandoning or supplementing quantum
theory}.
That means that in this context invoking of
``new physics'' is not needed; even more strongly, it is
counter-indicated.

A separate problem in the measurement process is the need to
describe how a result of an interaction between the measured
object and the apparatus on
the quantum level is transferred to the classical
level, e.g., to a pointer position.  We shall address both these
problems in our paper.

Quite generally, a theory is
supposed to be able to make ``predictions'' in the sense that given a ``state
of the system'' at some time, say, $t_0$, the theory must be capable of
providing information
on the state of the sytem at time $t$, where $t$ may be later or earlier
than $t_0$. The ``predictions'' here can apply before the relevant
experiment has been performed, or
after. Also, the other terms, e.g., ``the state of the system'', are
defined within the theory. In quantum physics, which is our framework,
this means (see Sections 7, 8) that given the
state of the system as $\rho(t_0)$ the probability of an outcome, ${\cal O}_k(t)$,
is given by $Tr\{U(t,t_0)~ \rho(t_0)~ U^\dagger(t,t_0) ~{\cal O}_k(t)\}$ where
$U(t,t_0)$ is the evolution operator of the system, and where $k$
specifies the outcome. For example, the outcome could be a particle
with spin-up emitted along a particular direction. In general there
are many, usually infinitely many, possible outcomes. However,
otherwise quantum physics is complete in that no hypotheses lying
outside of the quantum physics framework are needed, for any and all
circumstances.

Quantum physics does not make predictions on which of the possible
outcomes actually will occur. In particular, it does not predict
``when the event (e.g., the decay of the nucleus) actually will take
place.'' Such a prediction, being outside of quantum theory, would be
in conflict with the Heisenberg uncertainty relations. One of the
attempts to overcome this ``limitation of the theory'' is the
deBroglie-Bohm pilot wave hypothesis~\cite{7}. This attempt
supplements the wave function by a ``particle
function" $X(t)$, called by the authors a ``hidden variable."  Even
though outside of quantum mechanics, this work in fact provided the
stimulus for Bell to derive the famous Bell inequalities~\cite{5}
which allow for the distinction by experiment between at least a
large class of hidden-variable theories and quantum physics.  The
experiments by now have come out in favor of the quantum physics
predictions~\cite{asp}.

On the other hand, the theory is silent on the choice of the initial
conditions. In particular, one will have to chose $t_0$ sufficiently far
back so that the ``switching-on'' transient is contained within ones treatment. This is for example the case when investigating the unperturbed~\cite{NE} or
perturbed~\cite{Z} non-exponential part of the decay of unstable or
metastable states. Another well-known case is that of the correlations and
of the statistics
in high-intensity particle beams. The theory will provide answers for any
$\rho(t_0)$ and ${\cal O}_k(t)$  even if the initial conditions actually
are not, and can not be, realized in Nature. The reasons for this
impossibility may be obvious, or may be extremely subtle. A discussion
of this is given in Ref.~\cite{DISS}; it lies outside of the scope of
the present paper.

We have used the term ``outcome'' rather than the usually employed
term ``measurement''. The reason for this is that ``measurement''
seems to imply the participation of an observer. By using the term
``outcome'' we want to indicate that one is free to define the system as broad
or as narrow as one likes. Thus one may want to include the observer
in the system, which the theory and the formalism allows; but this
choice is irrelevant for our discussion.

In the present essay we want to show that Bell's conceptual gap~\cite{3}
disappears when recognizing that the particle-wave duality is only an
artifact of quantum mechanics, related to the absence there of the
particle aspects which are amputated when going from quantum field
theory to quantum mechanics.  The wave function contains only the
wave aspects.  Quantum fields (relativistic or non-relativistic)
contain simultaneously both the
particle and the wave aspects. The concepts required to construct
the needed framework are very few, and very simple. In the present
context the full interacting quantum field theory is not required;
one does not have to go beyond the lowest order, i.e., no Feynman
graphs containing loops need be considered.  This limited theory
is known to pose no mathematical difficulties~\cite{6}.  This then is the
subject we are going to address first (in Section 2).  Next, in
Section 3, we shall show in which way quantum mechanics (QM)
is a sub-field of quantum physics (QP), and then we will be in the
position to describe the process of measurement. We do that in the
next Sections by analyzing a series of experiments, and show that
both the wave and the particle aspects are needed in the description
of the measurement process.  As we will see, no conceptual gaps remain
when using both these aspects.  In particular, no split between the
quantum system and the conscious
experimentalist is needed; the experimental apparatus, including the
experimentalist, can be considered to be part of the quantum system.

Essentially {\it all of the concepts} needed in the description of the
measurement in quantum physics are present already in the case of a
\hbox{two-slit} experiment, Section 4, where the measuring arrangement
consists of an array of detectors.  The interference pattern which
arises in response to the wave aspect emerges as the result of a large
number of experiments, i.e., as a probability distribution.  The particle
aspect manifests itself in forbidding coincidences: in a weak beam
situation only one counter at a time can register an event.  Here
already the ``collapse of the wave function'' and ``communication at
faster than the speed of light'' between the  counters of the array
has to be operative if one wants to describe the situation in the
frame of quantum mechanics. The last of the concepts, the sensitivity
of the interference between different reaction channels to an intervening
measurement, here also can be fully elucidated: a measurement which
determines the slit through which the particle
travelled destroys the interference pattern.  Such a measurement
changes the \hbox{two-slit} into a \hbox{single-slit} pattern.

The Einstein-Podolski-Rosen (EPR) experiment~\cite{4}, discussed
in Section 5, is the generalization of a \hbox{two-slit} experiment
to a \hbox{two-particle} system; if one wants to describe it in
quantum mechanics one must replace the \hbox{3-dimensional} space
of the two-slit experiment by a \hbox{6-dimensional} configuration
space.  It requires a somewhat more complex experimental
arrangement and also a more complex theoretical analysis.  It is
the simplest setup allowing for \hbox{two-particle} coincidences.
When viewed within the frame of quantum mechanics in
\hbox{3-dimensional} position space it has indeed the well-known
dramatic conceptual problems.  Not so in quantum physics: the
description of all these experiments is fully contained within its framework.

In the next two Sections we discuss in detail the completion of
the quantum measurement, i.e., the reaction of a classical system
(a pointer, Schr\"odinger's cat) to the result in the quantum
system. This involves the description of a classical system in
terms of quantum physics; the mathematics needed for this
description is sketched in Appendix A.

No split between the
``quantum system'' and the ``classical apparatus'' is needed; all
has to be, and can be, considered from the quantum point of view.
Another frequently ignored aspect is that measurement inescapably is an irreversible
process, i.e., is associated with dissipation~\cite{DISS} taking
place already on the quantum level. This point is important and
must be taken into
account whenever a complete description, including the measuring
apparatus, is attempted.

The present paper does not address the question of the logical
superstructure, denoted ``the interpretation of quantum mechanics''
in Refs.~\cite{2,2'}.  The arguments and descriptions of
these references in fact involve assumptions concerning the
measurement process which are consistent with, actually follow
from, the results of the present paper.

In summary, all aspects required in the description of the evolution
of physical
systems, including the act of measurement, are contained
within quantum physics. Of course, the 19-th century
dream of a fully deterministic description, which would transcend the
frame of quantum physics, remains unfulfilled.
%\end{document}

\section{Quantum Physics}
In order to make the paper self-contained we now collect the
rudimentary aspects of quantum physics required for the present
purpose~\cite{6}.

In quantum physics the particular state under consideration is
described by a state vector, $|S\rangle$.  Thus, for example,
$|S(x_1,x_2,t)\rangle$ represents a state such that at time $t$
the system had two particles, one located at $x_1$, the other at
$x_2$.  The state
vector for the system which has no particles is given the special
notation $|V \rangle$, and is denoted as ``the vacuum.''  The field
operator, denoted by  $\Psi(x,t)$, {\it interrogates} the state vector
for the presence of a particle at the point $x,t$ in the form
\begin{equation}
\Psi(x,t) ~|S(y,t)\rangle ~=~ \delta(x~-~y)~|V\rangle  \label{I.1}
\end{equation}
with
\begin{equation}
\Psi(x,t) ~|~V\rangle ~=~ 0  \quad   \label{I.a}
\end{equation}
(in these equations, and throughout in this paper, we shall use
units such that $\hbar$ and $c = 1$).
Hence one calls $\Psi$  a ``particle annihilation operator.''
$\Psi(x,t)$ and $\bar\Psi(x,t)$ are defined to obey the
anti-commutation relations (commutation relations for Bosons)
\begin{equation}
\left[\Psi(y,t) ,~\bar\Psi(x,t)\right]_+~=~ \delta(x~-~y)~  \label{I.2}
\end{equation}
where the $\delta$-function implies the structure of a point
particle. Comparing (\ref{I.1}) and (\ref{I.2}) one sees that
\begin{equation}
|S(x,t)\rangle ~=~  \bar\Psi(x,t) ~|V\rangle  \quad . \label{I.3}
\end{equation}
In view of (\ref{I.3}) one calls $\bar\Psi(x,t)$ the ``creation
operator;'' at the same time one sees from (\ref{I.2}) that this
operator creates a point particle at the position $(x,t)$.  The field
operator $\Psi(x,t)$
is defined to obey the appropriate equations of motion, e.g., in the
nonrelativistic case the
Schr\"odinger wave equation
\begin{equation}
\left(i~\partial_t - H\right)\Psi(x,t)~=~0  \quad  .   \label{I.4}
\end{equation}
If relativistic effects are important, the Dirac equation or other
appropriate equations apply, in which case one has to consider the
negative-energy solutions and so on but our arguments below will not
be affected.

Equations~(\ref{I.2}) and~(\ref{I.4}) mean that $\Psi(x,t)$ has particle
character which propagates as a wave. In other words, the quantity
$\Psi(x,t)$ of quantum physics does not suffer from \hbox{particle-wave}
duality; it has simultaneously both particle and wave
characteristics.  This, of course, is not possible in classical
physics -- nor in quantum mechanics.

Both for computational purposes and for visualization, it is useful
to factorize $\Psi$ into the particle and the wave aspects.  This
can be done by computing a complete set of {\it c-number} functions, say
$\psi_n(x,t)$ which obey the wave equation (\ref{I.4}) {\em together
with the boundary conditions appropriate to the system}.  Then one
can write the expansions
\begin{equation}
\Psi(x,t)~=~ \sum_n~ b_n~\psi_n(x,t)     \label{I.5}
\end{equation}
for the field and
\begin{equation}
\bar\Psi(x,t)~=~ \sum_n~ b_n^\dagger~\bar\psi_n(x,t)
\label{I.6}
\end{equation}
for the hermitian conjugate field.  Inserting these definitions in
(\ref{I.2}) one sees that this equation is fulfilled if the
quantities $b_n,~ b_{n'}^\dagger$ obey the anti-commutation
relations
\begin{eqnarray}
\left[ b_n,~ b_{n'}^\dagger\right]_+ &=& \delta_{n,n'},     \label{I.7}\\
\left[ b_n^\dagger,~ b_{n'}^\dagger \right]_+ &=& 0~ =~ \left[ b_n,~ b_{n'}
\right]_+  \label{I.7a}
\end{eqnarray}
which then leads to the completeness relation in the form
\begin{equation}
\sum_n~\bar\psi_n(x,t)~\psi_n(y,t) ~=~ \delta(x~-~y)
\label{I.8}
\end{equation}
for the complete set of the solutions.
It is useful to introduce the abbreviation
\begin{equation}
\Psi_n(x,t)~=~ b_n~\psi_n(x,t) \quad .    \label{I.9}
\end{equation}
Herewith
\begin{equation}
\Psi(x,t)~=~  \sum_n~ \Psi_n(x,t) \quad .    \label{I.9a}
\end{equation}

The content of the Eq. (\ref{I.9}) can be
expressed as: acting on the vacuum the operator $\bar\Psi_n(x,t)$ through
the operator $b_{n}^\dagger$ creates a
particle in the
state $\psi_n(x,t)$.  Or, said
differently, $b_{n}$ is the particle aspect, and $\psi_n(x,t)$ is
the wave aspect of the quantum physics function $\Psi_n(x,t)$.  For
example, the anticommutation relations (\ref{I.7}) ensure that at most
one particle can occupy the state $\psi_n(x,t)$.  In the next Section we
will argue that $\psi_n(x,t)$ is linked to the probability interpretation of
quantum mechanics.

We complete this description by giving the expression for the
above-mentioned two-particle state vector
\begin{equation}
|S(x_1,x_2,t)\rangle ~=~ \bar\Psi(x_1,t)~\bar\Psi(x_2,t)~|V\rangle
\quad . \label{I.10}
\end{equation}
On the other hand the state vector of a system having a particle in the
state  $\psi_n(x,t)$  is
\begin{equation}
|S_n \rangle~=~ b_n^{\dagger}~|V \rangle  \quad . \label{I.10a}
\end{equation}
Below we will deal mostly with systems of that kind.

\section {Quantum Mechanics}
The basic concept of quantum mechanics is ``the wave function,''
also called ``the probability amplitude.''  The wave function for
the mode $n$, e.g., the state $n$ of the hydrogen atom, is denoted
as $w_n(x,t)$.  The meaning of this notation is defined as: given
any position $x$, and any time $t$, the value of the wave function
is the number $w_n$. The probability of  finding the particle there
is then $|w_n(x,t)|^2$.  To compute the wave function itself one
must solve the Schr\"odinger equation, or, if relativistic effects
are important, the Dirac equation, imposing the appropriate
boundary conditions on the solutions.  Both these equations are
wave equations.

How does this wave function emerge from quantum physics of the last
Section?  We make the ansatz
\begin{equation}
w_n(x,t)~=~ \langle V~|~\Psi(x,t)~|~S_n \rangle \quad ,
\label{II.1}
\end{equation}
where
\begin{equation}
|S_n\rangle~=~ b_n^\dagger~|V\rangle         \label{II.2}
\end{equation}
is the state vector for the system in state $n$.
Equations~(\ref{II.1}) together with (\ref{I.9}) yield
\begin{equation}
w_n(x,t)~=~\psi_n(x,t)        \label{II.3}
\end{equation}
which turns out to be consistent since both $w_n(x,t)$ and
$\psi_n(x,t)$ fulfill the same equation and the same boundary
conditions.  This shows that the quantum mechanics wave function is
the wave part of the quantum physics
function.  The function $w_n(x,t)$ describes  only ``the wave
aspects'' of quantum physics; {\it it lacks ``the particle
aspects''} which have been lost in the interrogation (\ref{II.1}).
Thus, in contrast to the quantum physics function $\Psi_n(x,t)$,
which describes a particle propagating, i.e., moving through space
and time, the quantum mechanics wave function $w_n(x,t)$ describes
a nothing propagating.  The latter is a rather abstract entity, having
led to many a fruitless search for the meaning of the
deBroglie-Schr\"odinger wave function, and, with it, the meaning of quantum
mechanics.  In order to have a description which contains {\it
both} the particle and the wave aspects one needs to work in
quantum physics.

\section {Preliminary Conclusions and Consolidation}
From our discussion above one sees that in quantum physics
complementarity, or, as it is also called, the particle-wave
duality, is necessarily absent since the state function~(\ref{I.9}),
$\Psi_n(x,t)$,
contains simultaneously both aspects.  It describes the motion,
i.e. the propagation, of a point particle through space and time.
This propagation is that of a wave, which precludes the possibility
of assigning a trajectory to that motion.  In contrast, quantum
mechanics simply lacks the particle concept, which is expressed by
the  interrogation formulae (\ref{I.1}), (\ref{I.a}), (\ref{I.2}).
The particle aspect has been eliminated from quantum mechanics at
the point where the wave function was extracted from the quantum
physics function in the interrogation (\ref{II.1}).  Thus, the
quantity which has been left intact, the wave function $w_n(x,t)$,
describes the propagation of nothing in particular, as  exemplified
by the Cheshire cat, which had left, leaving only the grin behind.

In short, only the wave function $w_n(x,t)$ exists in quantum
mechanics.  Of course, the wave function is an exceedingly rich
object, as can be seen from the scope of quantum mechanics.
However, the particle concept is inescapably needed for the
understanding, the interpretation, of the physical content of the
results obtained upon computation of the wave function.  It
therefore must be re-inserted artificially ``by hand.''  (In the
limited domain ``quantum mechanics'', which does not include, for
example, the radiative corrections, in the calculations themselves
the particle aspect is not needed.) This then leads to the particle-wave
duality, a logical gap, with the concomitant difficulty in
reaching full understanding.  This gap, which in fact is Bell's
above quoted ``problem of principle'', is one of the factors, most
likely the principal factor, which generated the aura of mystery and fog
%\cite{W}
surrounding the subject ``modern physics.''

\section{The Two-Slit Experiment}

Any measurement requires an interaction between ``the system'' and
the ``measuring device'', and in quantum physics every interaction
involves the emission or absorption of a particle (recall the
interaction term $\bar \psi \gamma^{\mu} A_{\mu} \psi$ of quantum
electrodynamics: a particle is absorbed in the initial state and is
re-emitted in a different, the final state; a photon is absorbed or
emitted). {\it Thus the measurement process lies outside
of the framework of quantum
mechanics.}  Let us discuss the process of measurement in terms of
specific examples. This present discussion will require a somewhat
more technical language than that of the previous
Sections.

Consider the determination of a diffraction pattern in a
photon \hbox{two-slit} experiment, and take the low-intensity case
to avoid the complication of chance coincidences.  The experimental
arrangement thus consists of a photon source, an intervening screen
with two (or one) slits, and an array of detectors behind the
slit-screen to register the photons and make the data available to
the experimentalist. The photon field, denoted here by $\varphi(x)$, can
be expanded in any set of  solutions of
Maxwells equations. Here it will be convenient to employ for the
two- and the one-slit cases two alternative such expansions, namely those
which obey the boundary conditions required
to account for the source, screen, two or one slits, etc. In the factorized
form of Eqs.~(\ref{I.5}),
(\ref{I.6}) the field then is given by (we change of notation from the
previous Sections
and suppress the vector character of the photon):
\begin{eqnarray}
\varphi(x)  ~=~ \sum_n~ a_n^{(2)}~ f_n^{(2)}(x)
~\equiv~ \sum_n~  \varphi_n^{(2)}(x)   \label{M.1}
\end{eqnarray}
if both slits are open, and
\begin{eqnarray}
\varphi(x) ~=~ \sum_n~ a_n^{(1)}~ f_n^{(1)}(x)
~\equiv~ \sum_n~  \varphi_n^{(1)}(x)      \label{M.2}
\end{eqnarray}
if only one slit is open. (Of course, the right-hand sides
of Eqs.~(\ref{M.1}) and (\ref{M.2}) can be expanded in terms of
either of $f_n^{(k)}(x)$. Expansion in the ``wrong'' function converges
uniformly
except in the vicinity of the slits.)  The individual terms of these two
sets of solutions,
$f_n^{(2)}(x)~,~f_n^{(1)}(x)~$, which actually are the wave functions
of quantum mechanics,
are different since
the boundary conditions for the two cases are different; in
particular, the interference patterns described by these two
solutions are different.  In these fields, $f_n^{(k)}(x)$ concerns
the wave aspects, while $a_n^{(k)}$ concerns the particle aspects:
$a_n^{(k)\dagger}$ creates, while $a_n^{(k)}$ annihilates, a
particle in state $f_n^{(k)}(x)$.
The two parts of the action of the detector, i.e., (i) the interaction
with the photons, and (ii) the registration of a ``count'' and the
transmission of the data to the user, and so on, factorize. The action
(i) of the detector $m$ tests for the presence of a particle by
interrogating the state vector at the space-time point $x_m$ (within
the resolution of the detector); it is described by the absorption
operator $\varphi(x_m)$,
both for the single-slit ($k=1$) or two-slit ($k=2$) case; see.
Eqs.~(\ref{M.1}),(\ref{M.2}).  Thus, for instance, the probability
amplitude for a particle in state $n$, being described by the state vector
$|S^{(k)}_n\rangle~=~ a^{(k)\dagger}_n|V\rangle$, to be at the point $x_m$, is
\begin{equation}
\langle V|\varphi^{(k)}(x_m)|S^{(k)}_n\rangle ~=~ f_n^{(k)}(x_m)~
\quad . \label{M.6}
\end{equation}
(The also correct form $\langle V|\varphi^{(1)}(x_m)|S^{(2)}_n
\rangle$ is inconvenient in that it results in a linear combination of
the functions $f^{(1)}_n(x_m)$.)
We collect the description of the action (ii) of the detector in an
appropriate operator $\eta$, which also includes the detector efficiency.
In this way, the action of the detector at the point $x_m$ can be described
by the detector function
\begin{equation}
D_m ~=~ ~ \sum_n \eta_m^n \varphi^{(k)}_n(x_m) \quad . \label{M.5}
\end{equation}
The extent of the sum over $n$ depends on the characteristics of the detector;
$\eta$ also contains the reaction of the measuring apparatus and hence is
an appropriate quantum operator.  The actual construction of the
detector is of no importance here; as an example, the
absorption of the photon may result in the ionization of an atom,
and the emitted electron may initiate a discharge as in a
proportional counter.

For a system having
an ``incoming'' photon in either the \hbox{two-slit} $(k=2)$ or the
\hbox{one-slit} $(k=1)$ situation,
the probability amplitude for the response of the detector
$m$ is
\begin{equation}
A_m^{(k)}~=~ \langle d_f |\otimes\langle V |D_m| S_n^{(k)}
\rangle\otimes | d_i\rangle
\sim ~\langle d_f| \eta_m^n |d_i\rangle ~f_n^{(k)}(x_m)
\quad ; \label{M.7}
\end{equation}
as to be expected the detector responds to the interference
pattern of the photon field, described by $f_n^{(k)}(x_m)$. Here
$\langle d_f|\eta |d_i\rangle$ denotes the expectation value describing the
detector response, from state $|d_i\rangle$ before the
interaction to $|d_f\rangle$ the state after detection.  As always,
the probability for counter $m$ to respond is $\left|A_m^{(k)}\right|^2$.

Both the particle and the wave aspect
contributed to the result, Eq.~(\ref{M.7}).  The particle aspect, the
factor $\varphi(x_m)$ contained in $D_m$, absorbed/annihilated the photon, and
this took place in a local manner,
precisely at the \hbox{(four-)point} $x_m$ in the detector
$m$; it also provided  the factor $f_n^{(k)}(x)$, i.e., the wave aspect,
which is the appropriate
solution to Maxwells equations.

Furthermore, once one detector has registered a photon, then no
other detector can respond since the particle already has been
absorbed.   This follows from the expression for the
probability amplitude, say $A_c$, of a coincidence in detectors $m$
and $m'$ (dropping the irrelevant factors),
\begin{equation}
A_c~\sim~ \left \langle V~|D_{m'}~D_m|~S_n^{(k)} \right\rangle
\sim ~ \left\langle V~|D_{m'}~|~V \right\rangle ~=~ 0 \quad ;
\label{M8a}
\end{equation}
the probability for a coincidence thus vanishes in view of Eq.~(\ref{I.a}). Expressed in spoken language, the meaning of Eq.~(\ref{M8a}) is: ``The anwer to the question `can two detectors, $D_m$ and $D_{m'}$, absorb a single particle, $S_n^{(k)}$?' is NO.'' The absence of a
coincidence is enforced by the particle aspect.

The requirement for a ``decision'' of hitting this one, or that one, but then no
other detector, is the {\it prototype of the need for the ``collapse
of the wave
function'' of quantum mechanics, which, of course, is not described
by the equations of motion,} e.g., the Schr\"odinger equation.
Namely, in quantum mechanics the counter $m'$ must somehow be made
to ``know'' that the counter $m$ has been hit.  The wave function
originally in general is \hbox{non-zero} at both places.
Thus there is no reason for counter $m'$ not to respond at the same
time. To avoid such a coincidence one therefore in quantum
mechanics must {\it mimic the uniquely local character} of the absorption
process. This is accomplished by ``collapsing the wave function'';
essentially from $f_n^{(k)}(x_{m})$ to a delta-function at  $x_m$
-- or at $x_{m'}$ if it was detector $m'$ which had responded.
This ``collapse'' is even more spectacular in the case of a more
complicated reaction where the wave function extends not only over
a small region of space but over a perhaps large number of reaction
channels; to accomplish this feat ``the needed  signal: `collapse
the wave function!' may have to propagate faster than light''. (Of
course, no such signal is available in quantum theory.) We
will return to this point below in the discussion
of the Einstein-Podolski-Rosen experiment.

In case one tries to check ``which slit the photon passed through''
one has to place a detector in the slit, say at $x_s$.  To know
that the photon passed through this slit this detector would have
to record a Compton scattering event.  In this detection process
{\it the original photon is absorbed and a new photon is emitted},
having a new energy and a new radiation pattern appropriate to the
new geometry (as required by the dependence of the solution on the
boundary conditions), i.e., radiation from within the slit.
The Compton detector function then would have the form
\begin{equation}
D_s ~\sim~ \eta_s ~\varphi^{(-)(s)}(x_s) ~\varphi^{(k)}(x_s) ~
\label{M.8}
\end{equation}
where $\varphi^{(-)(s)}(x_s)$ is the emision part of
$\varphi(x_s)$ for a photon with
the new radiation pattern (that of only one slit open) constructed
in analogy to
Eqs.~(\ref{M.1}) or~(\ref{M.2}).  $\eta_s$ is as previously the
operator describing the reaction of the detector, here the
recoiling atom.  The new radiation pattern is different
from the old radiation pattern.  In particular, it does not exhibit
the two-slit interference pattern.

The two-slit interference pattern will
disappear upon the Compton scattering of the photon {\it even if nobody
actually observes the counter}, or even if the counter is broken.
Such processes, of course, take place all the time; they are called
``collimator scattering'' and contribute to the experimental
background.  That means that the photon needs not ``to know that it
has been observed'' to change the  interference pattern.  This way
Nature in quantum
physics ``has an objective existence; it exists by itself''
independent of observation.  Hence the experimental apparatus, and
by extension the experimenter, can be considered to be ``part of
the system'', without any difficulties, conceptual or otherwise.

Instead of putting a detector in one of the slits to find out which
slit the particle passed through, one may place a screen
over one of the slits after the particle has been emitted. This
case requires the description by localized wave-packet states which will be
introduced in the next Section. Then, independently
of whatever happened before or after, the state of the system is
given by the solution of Maxwells equations appropriate to the
boundary conditions valid at the time when the wave packet
arrives at and passes through the slits. The state of the system at
the time when the photon just has been emitted is given as always by
\begin{equation}
\varphi(x,t) ~=~
\sum_n\,'~C_n^{(2)}~ f_n^{(2)}(x)~ \left( a_n^{(2)}~
e^{-iEt}\right)
~+~ \sum_n\,'~C_n^{(1)}~f_n^{(1)}( x)~ \left( a_n^{(1)}~
e^{-iEt} \right)
\end{equation}
where the coefficients $C_n^{(2)}$ and $C_n^{(1)}$ are determined
by the emission process. The primes at the summation signs indicate
that the sums are to be taken over the functions appropriate to the
case to avoid double counting. (If the screen with the slits is ``very
far'' from the source, then $C_n^{(1)}$ and $C_n^{(2)}$ may be
equal.) The probability for registering a count therefore will
contain the factor $ \left|C_n^{(k)} \right|^2$,
multiplying the square of the right-hand side of Eq.~(\ref{M.7}).
Thus, in all of these experiments, the existence or non-existence
of a diffraction pattern or a coincidence is determined by the
particle aspects while the probabilities are given by the wave
aspects; and no collapse
takes place or is needed.  In other words, the ``yes or no'' is
determined by the particle aspects, the ``how much'' by the wave
aspects of quantum physics.

\section{The E-P-R Experiment}
Particularly notorious as a ``paradox'' is the
\hbox{Einstein-Podolski-Rosen} example~\cite{5,4} which
in quantum mechanics
for its ``explanation'' combines the need for wave function collapse
and action at a
distance.  With respect to the above
\hbox{two-slit} example, the EPR setup has two added features
(which lead to complications of detail, but not  of principle): (i)
the system contains two particles; and (ii) the time-dependence
must be accounted for.

Consider the point (ii) first.  An eigenstate of the Hamiltonian by
definition has a precise energy.  Hence the wave function of an eigenstate
factorizes as
\begin{equation}
\phi(x,t) ~=~ e^{-iEt}~ \phi(x) \quad . \label{M.18}
\end{equation}
The probability to find the particle at the point $x,~ |\phi(x,t)|^2 =
|\phi(x)|^2$, hence is independent of the time.  With this solution
thus it is impossible to specify a time as being before, during, or
after the experiment.  To achieve this possibility one must chose
a suitable superposition of these states.  We follow the Weyl
prescription.  Thus for a free particle we chose the form
\begin{equation}
w(E_k;x,t) ~=~ {\cal N} ~ \int_{E_k-\Delta}^{E_k+\Delta}~
e^{i[px-E'(t-t_0)]}~ dE'  \label{M.19}
\end{equation}
where ${\cal N}$ is the normalization constant.  This function is
localized: it peaks at $x = 0$ at the time $t = t_0$; as $t$
increases the peak propagates towards larger $x$ with a velocity
the corresponding classical particle would have.  As is well known,
one must distinguish between the phase velocity, $a = E/p$ and the
group velocity, $b = dE/dp$.  It is the latter which corresponds to
the classical particle velocity; in view of
\begin{equation}
p ~=~ v~ \sqrt{ m^2 + p^2}        \label{M.19b}
\end{equation}
and
\begin{equation}
E ~=~ \sqrt{ m^2 + p^2}  .       \label{M.19c}
\end{equation}
we find $b = v$. On the other hand $a = 1/v$. The width $\Delta$ of
the superposition determines the width of the peak in $x$;
also, $w(E_k;x,t_0)$ is a minimum-uncertainty
wave function.

With functions of this kind one thus can describe,
for example, the following: the particle was emitted from the
source at time $t_0$, arrived at the scattering center at time
$t_1$, and was absorbed in the detector at time $t_2$.  From now on
we will work exclusively with Weyl functions.

We explain the point (i) above directly in terms of the EPR
setup. There at time $t=t_0$ two particles of spin $s = 1/2$,
coupled to total spin $S = 0$, are emitted in opposite directions,
go through a series of polarizers and analyzers to be finally
absorbed in two widely separated detectors.  The point of the
experiment consists in changing the setup at random after the
particles have been emitted and have become separated by such a
distance that ``they cannot communicate'' without violating
relativistic causality.

We need the following definitions. As the functions $w$, Eq.~(\ref{M.19}),
form
a complete set the field operator can be expanded as
\begin{equation}
\varphi(x,t)~=~\sum~a_n~w_n(x,t) \quad . \label{m20a}
\end{equation}
Since we will only deal with the system when the particles
are far from the source it is convenient to split the field into
separate field operators for the two directions.  Denoting spin ``up'' and
``down'' by the indices $+$ and $-$ respectively, taking the
quantization to be along the z-direction, we have for the
split field operators (suppressing the time)
\begin{equation}
\varphi_w(x) ~=~ \sum_n~( a_{n+} ~ w_{n+}(x) ~ +~  a_{n-} ~ w_{n-}(x)),
\quad x~>~0
\quad ;  \label{M.20}
\end{equation}
\begin{equation}
\varphi_v(x) ~=~ \sum_n~( b_{n+} ~v_{n+}(x) ~-~  b_{n-} ~ v_{n-}(x)),
\quad x~<~0 \quad . \label{M.20A}
\end{equation}
Thus here $w$ denotes a particle emitted ``to the right,'' and $v$ ``to
the left''.  Since the particles
are described by Weyl functions they indeed fly apart. The state vector
of the system then is
\begin{equation}
|S_2 \rangle ~=~{\cal N}~(a_{n+}^{\dagger}~b_{n-}^{\dagger}~-~
a_{n-}^{\dagger}~b_{n+}^{\dagger}) ~ |V\rangle   \quad ,  \label{M.21}
\end{equation}
with ${\cal N}$ the normalization factor; note the absence of the summation
over $n$.  In the following we will suppress the quantum number $n$ for
brevity.

The detectors are endowed with polarization analyzers, and are
placed at $-X$ and $+X$ respectively with ``very large'' separation $2|X|$.
Denoting the polarization of the analyzer by the subscript $p~=~+$ or $-$ the
detector response is given as
\begin{equation}
D_1 ~=~ a_{+}~w_+(x)~ \eta_+~ +~ a_-~w_-(x)~ \eta_-  \label{M.22}
\end{equation}
and
\begin{equation}
D_2 ~=~ b_{+}~v_+(x)~ \eta'_+~ +~ b_-~v_-(x)~ \eta'_- \label{M.22a}
\end{equation}
where detector 1 is at $x=X$ and detector 2 at
$x=-X$.  The probability for obtaining a coincidence, i.e., a count
in both detectors, irrespective of whether the spin state is ``up'' or
``down' then is given by (we suppress the detector states)
\begin{equation}
A ~\sim~ \langle~ V | ~ D_1 ~ D_2 ~ |S_2~\rangle \quad .
\label{M.23}
\end{equation}
The quantum probability for detection thus will have an interference
term for spins at $X$ opposite to that at $-X$, as can be readily
deduced from~(\ref{M.22}).

Now insert a polarization-sensitive filter
in arm 1, such that only the ``up'' state is transmitted.  This
filter is represented by the projection operator
\begin{equation}
F ~=~ a_+^\dagger ~ a_+~    \label{M.24}
\end{equation}
which has the effect given by
\begin{eqnarray}
F~ a_+^\dagger ~ |V\rangle ~=~ a_+^\dagger ~ a_+~  a_+^\dagger ~
|V\rangle ~=~ a_+^\dagger ~ |V\rangle
\end{eqnarray}
\begin{equation}
F~ a_-^\dagger~|V\rangle ~=~ a_+^\dagger~ a_+~ a_-^\dagger~|V\rangle~
=~ 0  \quad .    \label{M.24b}
\end{equation}
Thus Eq.~(\ref{M.23}) is replaced by
\begin{equation}
A ~\sim~ \langle V | ~ D_1 ~ D_2 ~F~ |S_2~\rangle \quad .
\label{M.25}
\end{equation}
Now only the term with spin ``up'' at $X$ and ``down'' at $-X$ survives.
That means, that in
a coincidence the detector in arm 1 ``determines'' the polarization
of the particle in arm 2.  And it does not matter at what time the
filter was inserted in the beam path, as long as that took place
before the arrival of the Weyl wave packet.  All these results
emerge directly in terms of the quantum physics functions; no
collapse of a wave function, or transmission of a signal, is
required.  To guarantee that, where appropriate, coincidences
indeed do occur requires that the detector efficiencies be 100\%.

A similar analysis can be carried out for the case of a spin-flip
filter,
\begin{equation}
T ~=~a_+^\dagger ~a_- ~+~ a_-^\dagger ~a_+ \quad , \label{M.26}
\end{equation}
the action of which is given by
\begin{equation}
T~ a_+^\dagger ~ |V\rangle ~=~ a_-^\dagger ~ |V\rangle \quad
\label{M.26a}
\end{equation}
or
\begin{equation}
T~a_-^\dagger~|V\rangle ~=~ a_+^\dagger ~|V\rangle   \quad .
\label{M.26b}
\end{equation}
Replacing in (\ref{M.25}) $F$ by $T$ of (\ref{M.26}), one finds
that coincidence is achieved when both detected spins are of the same
``orientation''.
As above, it does
not matter when the filter was inserted. Again, the counter in
one of the arms ``determines" the polarization of the particle in
the other arm.  The importance of coupling to spin $S = 0$ is
manifested by the minus sign in Eq.~(\ref{M.21}).

As there is no privileged direction, one can choose the polarization
detectors to be sensitive along the y-direction.
Then the detector
response function~(\ref{M.22}) is represented by $a'_{\pm}$ and $b'_{\pm}$,
the annihilation operators quantized along y-orientation.  The state vector
of the
system still can be represented by~(\ref{M.21}). However, it now is more
convenient
to expand the field operator~(\ref{M.20}) in the basis of y-oriented
wave functions, and
the above analysis will go through with ``up'' and ``down'' now referring
to the
y-axis.  Whatever the representation of the field operator, we obtain the
same
result.

An interesting case arises when the analyzer of the detector in arm 1
measures the spin
along the z-axis and that of arm 2 along the y-axis, say.
The state then would appear to the detector in arm 2 as
having terms not only of the form $a^\dagger_+~b'^\dagger_-$ but also of
the form $a^\dagger_+~b'^\dagger_+$. Therefore no strict yes-no
coincidence
rules would exist and only probability predictions would be possible.
It is precisely these probabilities which are different in quantum
physics and in classical probability.  The analysis of this
situation forms the basis for the Bell inequalities.  All of the
Bell predictions, Ref.~\cite{5}, made in the framework of quantum
mechanics are correct, i.e., are in full agreement with those
derivable from the above quantum
physics description -- except
for requiring an acausal propagation of the hypothetical signal
``inducing the collapse of the wave function''.

\section{Dissipation and Decoherence in Measurement}

Above we demonstrated that the measurement process in
quantum physics poses no conceptual problems when describing it in terms
of quantum fields. In that discussion we paid no
attention to the full measurement process, which actually can be split
into two interdependent stages: the interaction with the quantum object,
and the chain of amplification to the macroscopic ``pointer position''.
The essential point in the second stage is the transfer of the results
of the measurement from the quantum physics (QP) part to the Classical
Physics (CP) part
of the apparatus.

So far we have concentrated only on the first of these
stages, i.e., the very first interaction between the
``system'' and the ``apparatus''.  In the present Section we shall develop
a more complete description, which will allow us to trace the process
through to the completion of the second stage, i.e., from the beginning
of the amplification chain to the display of the result of the measurement
by a ``pointer position''.  Our aim is to derive the expression for the
probability, say $P_p$, of the pointer position $p$ after the conclusion
of the measurement process. In essence this amounts to the full treatment
of the matrix element $\langle d_f|\eta|d_i \rangle$ of Eq.~(\ref{M.7})
postponed in Section 5.

Since the ``yes or no'' question has been answered in the first of the
two stages of the overall measurement process, the second part which
concerns the ``how much'' aspect could be discussed simply in the
quantum mechanics (QM)
framework. For completeness we will, however, also discuss it in the
full QP frame.

The essential new aspects to be treated now will be: (i) the inclusion
of dissipation which is inherent and inevitable in any actual measurement
since any elementary quantum act of the measurement process itself is
irreversible, i.e.,
dissipative~\cite{DISS}; and (ii) the process of decoherence, which,
as we will see, is the physical basis for the conversion from a quantum
to a classical process~\cite{FRMP,DGD}. Superficially, these points seem
similar but we will see that actually they differ in an essential manner.
Of these two aspects, in particular
(ii) requires that the description be done in terms of density matrices
(see Appendix A), needed when
dealing with non-interfering, ``classical'' objects. To give a full
description we will have to assemble some needed QP quantities, in
particular in the context of decoherence.

Concerning point (i), the very interaction underlying the measurement
involves a time-reversal non-invariant evolution of the system
object-plus-device, no matter whether the objects are quantum or
classical. The dissipation in the macroscopic, ``classical'',
down-part of the chain of the overall process poses no problem.
Our subject here thus is the analysis of
the dissipation associated with the elementary, the quantum measurement
process at the very beginning of the chain.

The point (ii) deals with the transition from the quantum to the classical
part of the chain. This involves the description of classical objects in
the QP frame.  It is essential in resolving the Schr\"odinger cat paradox.

Begin with point (i), the initial interaction. For definitiveness we
take as the example of the initial ``quantum detector'' a Compton
proportional counter as used
to determine ``the slit the photon went through'' in the two-slit
experiment of Section 5. Such a counter is filled with a low-pressure
gas and contains the needed electrodes.  The photon may suffer Compton
scattering on one of the electrons belonging to one of the gas
molecules, the gas being a classical object of given temperature
and pressure. We thus need the quantum description of that classical
molecule, both of its motion and of its internal state,
for example its rotation-vibration quantum numbers, and so on.

Consider the translational motion. Being confined to the counter
volume, it can not be described by a plane wave. Instead, the basis
states can be taken as Weyl packets, Eq.~(\ref{M.19}). Since it is
a classical non-interfering state, it must be described by a density
matrix (see Appendix A). That means, the density matrix must be made
up of Weyl basis states, each of which is confined to the counter
volume. Denoting the energy expectation of
a Weyl basis state, $\phi_i$, by $E_i$, i.e.,
\begin{equation}
E_i~=~\langle \phi_i~|~H~|~\phi_i \rangle \quad , \label{DM1}
\end{equation}
the elements of the density matrix will be
\begin{equation}
\omega_{ij}~=~ {\it Z} \delta_{i,j} e^{-E_i/T} \quad , \label{DM2}
\end{equation}
with ${\it Z}$ the normalization constant, and $T$ the temperature.
In (\ref{DM1}) the index $i$ may encompass the internal quantum
numbers of the molecule. (For room temperature and taking the gas
of the counter to be argon we obtain the localization of the atoms,
i.e., their position
uncertainty, to be less than atomic dimensions, thus
negligibly small.) The form of the density matrix for an individual
molecule thus is of block-diagonal form: each Weyl basis state,
being a wave function as in Eq.~(\ref{M.19}), is given by a block
of fully interfering components, while no interferences exist
between different blocks. The density matrix of the complete
system, of the proportional counter, is given by the direct
product of the individual-molecule density matrices.

To continue we need the QP expressions (see Appendix A)
corresponding to the
QM expressions used above in Eqs.~(\ref{M.19}), (\ref{DM1}), (\ref{DM2}).
To that end we write the expansion of the
Weyl states in the usual plane wave basis $\chi_i$:
\begin{equation}
\phi_i~=~\int dp g_i(p) \chi_i(p)  \quad , \label{DM3}
\end{equation}
where the functions $g_i(p)$ form a suitable complete orthonormal
set chosen in particular to ensure the fulfillment of the
boundary conditions, i.e., vanishing outside of the counter. (The form
of Eq.~(\ref{DM3}) is more general than that of Eq.~(\ref{M.19}).) Then
the QP field operator is written~\cite{DGC}
\begin{equation}
\psi_i~=~\sum_i~B_i \phi_i  \quad , \label{DM4}
\end{equation}
where $B_i$ is the annihilation operator for the Weyl mode $i$,
\begin{equation}
B_i~=~\int dp g_i(p) b_i(p)  \quad . \label{DM5}
\end{equation}
Herewith we can write the (bilinear) state vector for the initial-state
density matrix
\begin{eqnarray}
{\widetilde {\it S}}^{(0)}~=~ \sum_{ij}~ B_i^{\dagger} |V
\rangle~\omega_{ij}^{(0)}~\langle V| B_j ~\equiv~ \sum_{ij}~{\widetilde
{\it S}_{ij}}^{(0)}
 \quad . \label{DM6}
\end{eqnarray}
We shall call these quantities ``density state vectors.''
Similarly we have {\it mutatis mutandis} for the final state
\begin{eqnarray}
{\widetilde {\it S}}^{(f)}~=~ \sum_{ij}~ B_i^{\dagger} |V
\rangle~\omega_{ij}^{(f)}~\langle V| B_j ~\equiv~ \sum_{ij}~{\widetilde
{\it S}_{ij}}^{(f)}
\quad , \label{DM9}
\end{eqnarray}
which we will have to compute by considering the evolution of the system.

Denoting the photon Fock-space operators by $a^{(0)},~a^{(f)}$ for the
initial and final photon state,
respectively, we have for the complete density state vector
\begin{eqnarray}
S^{(k)}~=~ \sum_{mn,ij}~a_m^{(k)\dagger}~\sigma^{(k)}_{mn}~
{\widetilde {\it S}}_{ij}^{(k)} ~ a_n^{(k)}~\equiv ~\sum_{mn,ij}
S_{mn,ij}^{(k)} \label{DM8a}
\end{eqnarray}
with $(k) = (0),(f)$, and $\sigma^{(k)}_{mn}$ the photon density matrix.

The interaction with the incoming photon is, as always,
\begin{equation}
H(x)~=~ie~\bar \psi(x) \gamma_{\mu} \psi(x) A^{\mu}(x) \quad ;
\label{DM7}
\end{equation}
here $\psi(x)$ can be expanded in terms of the functions $\psi_i$,
Eq.~(\ref{DM4}). As we are interested in a Compton process, the
full interaction, including the rest of the apparatus, indicated
as previously by $\eta$, is
\begin{equation}
I~\eta~=~ H(y)~G(y,x)~H(x)~\eta  \quad . \label{DM8}
\end{equation}
Here $G$ is the relevant Green function, describing the propagation
of the electron from $x$ to $y$.

Concerning dissipation in the measurement interaction induced by
Eq.~(\ref{DM8}), every initial
state, i.e., every incoming reaction channel, leads to several,
actually to a very large number, of reaction channels. This is the
signature of dissipation, as shown in Ref.~\cite{DISS}, and as is
immediately visible from the quantum Boltzmann formula for the
entropy $\cal E$~\cite{DISS}:
\begin{equation}
{\cal E}^{(k)}~=~ - \sum_j~ \omega_{jj}^{(k)}~ log~ \omega_{jj}^{(k)}
\label{DM11}
\end{equation}
where the superscript $(k)$ indicates the initial or final state.
There are very many more states, $j$, in the final state than in
the initial: the entropy has increased in the measurement interaction.
It does not matter whether the final-state degrees of freedom
are observed or not; they participate in the entropy Eq.~(\ref{DM11}).
(See Ref.~\cite{DISS} concerning irreversibility.)

Now to the point (ii) above, the question of decoherence. That process
arises when the full density matrix is replaced by the ``effective''
density matrix, i.e., that part of the density matrix associated with
the relevant degrees of freedom, and when in that replacement the density
matrix loses some or all of its off-diagonal elements by tracing over the
unobserved degrees of freedom -- {\it which, in fact, is the only possible
mechanism for decoherence}~\cite{FRMP}. (This tracing has the
same origin as the integration over the unobserved degrees of freedom
when contracting a fully differential cross section to a partially
differential cross section.) This way we expect
that for the final state, being a classical system, the density
matrix $\omega_{ij}^{(f)}$ also will be diagonal, i.e., fully
decohered. However, since the
incoming photon may be in a pure state, that expectation may not be
fulfilled; this has to be investigated.

The complete response of the system, the probability $P_p$, then is
given by the general expression
\begin{equation}
P_p~=~ Tr' \langle~I^{\dagger} {\it S}^{(0)}~I~ {\it R} ~\rangle
\quad . \label{DM10a}
\end{equation}
The structure of (\ref{DM10a}) is: the state after the initial
interaction, $I^{\dagger} {\it S}^{(0)}~I$, undergoes evolution by the
action of the system evolution operator $R$, associated with the
operator $\eta$. We now must decide how far down the chain of evolution
we want to follow. Thus one may include the observer in the description;
this would require knowledge of the QP mechanism of the brain action,
which we do not possess. We shall, however follow the chain in
principle -- even though not in detail -- down to the ``pointer
position''. As we will see, it is fully sufficient to break off the QP
chain much earlier, essentially immediately after the elementary act
of the quantum measurement, and continue its description as a CP object.

In Eq.~(\ref{DM10a}) $R$ is written in the density matrix form. Hence,
in terms or the Tomonaga -- Schwinger evolution operator, i.e.,
the $U$-matrix~\cite{6}, $R$ is of the form
\begin{equation}
R~=~ U(t,t_0)~ U^{\dagger}(t,t_0) \label{DM10b}
\end{equation}
which is the full QP expression. It gives the state of the system at
time $t$, i.e., the fully differential transition probability into
all channels. To obtain the probability for a particular final state,
say, a given pointer position, one sums over all unobserved degrees
of freedom, i.e., one computes the full trace leaving out the observed
variables -- hence the prime at the trace symbol.  Eq.~(\ref{DM10a})
is a QP expression.
%To arrive at the probability, $P_p$, one must
%evaluate the Fock matrix elements in analogy to %Eq.~(\ref{M.7}). That
%is indicated by the symbols $\langle~,~\rangle$ in %Eq.~(\ref{DM10a}).
%In the following we shall omit these symbols. This way the %expressions
%will have the form of QM expressions; one will, however, %where needed
%have to provide explicitly the ``yes or no'' of the implied %QP to
%achieve the correct description.

We recognize that, in the notation
of Section 5, the counter $m$, Eq.~(\ref{M.5}) will respond for all
(photon) states $\varphi_n(x_m)$ which are non-zero at $x_m$. The
different states $n$ are here associated with different recoils of
the atom having undergone the Compton scattering. They provide the
initial states for the evolution operator $R$, Eq.~(\ref{DM10b}),
i.e., the sate of the system at time $t_0$. The operator $R$ contains
the downstream dynamics, in particular the (hypothetical!) recoil
detector. Even though in our example of the Compton counter such a
detector can not be built, in simpler cases it might be possible to
achieve the appropriate measurement \cite{PRL7}. Then in this step no
decoherence
would occur; the interference terms between the reaction channels would
be retained.  Thus, in contrast to dissipation, which is inevitable,
{\it decoherence is not a basic law of nature}.

We now discuss the decoherence in some more detail, considering a
system of adequate complexity to qualify as a (potentially) classical
apparatus. As the system
evolves down the amplification chain it not necessarily loses
coherence immediately. That can be seen as follows.

Taking the sytem at the beginning of the step $c$ of the chain to be in a
pure state, i.e., to be given by a wave function (for brevity we write
here in analogy to Eq.~(\ref{II.1}) only the wave function part
remaining after the evaluation of the
Fock operator matrix elements)
\begin{equation}
\psi~=~ \sum C_i~\varphi_i \label{dc1}
\end{equation}
then, at the next step, $c'=c+1$,
\begin{equation}
\psi'~=~ \sum C'_j~\varphi_j \label{dc2}
\end{equation}
and similarly with $c''=c+2$,
\begin{equation}
\psi''~=~ \sum C''_k~\varphi_k  \quad . \label{dc3}
\end{equation}
The relation between the consecutive density matrices in terms of
evolution operator $U$ is
\begin{eqnarray}
\rho''_{kk'}~&=&~C''_k~C''^*_{k'}\cr
\cr
&=&~ \sum_{j,j'}~U_{kj}~U^*_{k'j'}~C'_j~C'^*_{j'}~=~ \sum_{j,j'}~
U_{kj}~U^*_{k'j'}~\rho'_{jj'} \cr
\cr
&=&~  \sum_{j,j';i,i'}~U_{kj}~U^*_{k'j'}~
U_{ji}~U^*_{j'i'}~C_i~C^*_{i'} \cr
\cr
&=&~\sum_{j,j';i,i'}~U_{kj}~U^*_{k'j'}~U_{ji}~U^*_{j',i'}~\rho_{ii'}
\end{eqnarray}
Here we have used the group character if the evolution operator
\begin{equation}
U(t_n,t_0)~=~\prod_{k=0}^{n-1}~U(t_{k+1},t_k)   \label{FF}
\end{equation}
and have introduced the compact notation $U_{j,k}=U(t_j,t_k)$.
The product of the U-matrices, upon continuation to the end of the
process, and evaluating the Fock operators, becomes the matrix of the
above operator $R$.

In a complex system the reaction channels tend to decouple. That means
that such a decoupled state, say $k$, is connected with the initial
state $i$ by a single chain of intermediate states, i.e., by a chain
without loops~\cite{DGD}. If that state is not one of the actually
observed states the density matrix at that point loses its relevant
off-diagonal matrix elements, as discussed in detail in the next
Section after Eq.~(\ref{C5}). Hence the up-stream density matrices
of the side chain in effect also lose their off-diagonal elements,
and one may break off the product of the U-matrices at the
point where this chain branches off, say at the state $j=j_0$, and
replace
the follow-up chain by the Kronecker $\delta_{j,j'}$, which means
short-circuiting the tracing over the states at time $t$ and following
all the branches back to the state $j_0$. In fact, in this case the
intermediate
density matrix, $\rho'$, in the row and the column $j_0$  then anyway
in effect has vanishing elements, except the diagonal, which has the
value $|C'_{j_0}|^2$. This thus will lead to reduction of the off-diagonal
elements of the density matrix $\rho''$, without eliminating all of
them. (Of course, the form $C''_k~C''^*_{k'}$
for  $\rho''_{kk'}$ then is not valid.)  In this case the de-coherence
is incomplete. Complete decoherence would result if all branches lead
to unobservable, or, at least, unobserved output channels.

This formulation is fully general. Thus, for example, in the
description of the quantum Zeno effect, Ref.~\cite{Z}, the
``probe pulse'' checking for the state of the system, could be
described in the evolution from $\rho$ to $\rho'$, i.e.,
incorporated in the evolution operator $U'$. Similarly, an
event in the history of the evolution, discussed in
Refs.~\cite{2} and~\cite{2'}, is based on the factoring Eq.~(\ref{FF})
and would properly be described in
terms of the appropriate operators $U$.

Returning to our discussion, in the further evolution of the
system in the next steps similar
partial decoherence effects can take place. As decoherence is
irreversible it is cummulative, and after a time it will
be essentially complete. In the case of our proportional counter this
full decoherence will need only very few steps, most probably
only one. Further details concerning the mathematics of decoherence
are given in Appendix B.

In summary, in any realistic
situation there exist many degrees of freedom which participate in
the measurement chain, and which are not being observed. In
our example of the Compton process such is the recoiling ion, and
the ionizations and delta rays arising from interactions of the
recoiling Compton electron with the gas.  No measurement of their
characteristics is contemplated or possible.  All they do is trigger
the discharge in the counter.  Hence here $ R_{pq}$ is diagonal,
i.e., it vanishes for $p \ne q$. That then eliminates any possibility of
interference in the final state~\cite{FRMP}; the decoherence is complete.

Once the discharge in the counter has taken place, quantum effects
become irrelevant, since in a decohered state coherence can not be
restored. Furthermore, the amplification chain and what
not is made of classical
apparatus. From then on in the quantum description only mixed-state
density
matrices participate, exhibiting no interferences between the
different possible outcomes.

\section{The Schr\"odinger Cat as a Pointer}
Even though the results of the previous Section implicitly provide the
complete answer to the problem of the pointer, known as the Schr\"odinger
cat, in view of the extensive discussions in the literature
we shall give a rather detailed discussion of this case.

The experimental setup consists of a quantum measurement
device, leading through a classical amplification chain to a two-position
pointer indicating whether the quantum event has or has not taken place.
Specifically, the pointer is taken to be a cat, and the two pointer
positions are represented by the cat being alive or dead: the quantum
event is amplified into killing of the cat. This setup thus is a particular
realization of the generic quantum measuring apparatus. The question posed
in this context is: Can one ``know'' whether the cat is alive or dead
before looking in the box? More specifically: Is the cat dead or alive
before (one of us!) looking? {\it These questions have the same
meaning, and permit the same answers, in both the classical
and the quantum contexts}.

The only question which is possible in the quantum and not in the classical
context is: Before we look, is the cat in the linear superposition
\begin{equation}
\Psi_{cat}~=~a(t)~\psi_{live}~+~b(t)~\psi_{dead} \quad ? \label{C1}
\end{equation}
Can one observe interference between $\psi_{live}$ and $\psi_{dead}$?
Does this wave function collapse into either ``live'' or ``dead''
upon our taking a look?

Recall the way interferences arise in experimental
situations, as for example the interference between electric quadrupole
and magnetic dipole reaction channels in photon absorption. The
state {\em after the
transition}, if pure, is
\begin{equation}
\Psi~=~\sum_{n,l}~a_{nl}~\psi_{nl}  \label{C2}
\end{equation}
with
\begin{equation}
\psi_{nl}~=~\chi_n~ \phi_l(\theta,\varphi)  \label{C2a}
\end{equation}
where $\phi_l$ describes the emitted particle with $\theta,\varphi$
the direction of emission,
and with $\chi_n$ describing the in general unobserved recoiling particle.

Assume that
the incoming photon is polarized with polarization $p$, and denote
the photon density matrix by $\sigma_{p'p''}$. Further assume that the
final state of the recoil reached in the transition (transition operator
denoted $T$) depend on the polarization.  Thus the state index $n$ splits
into two classes, say $u$ for the ``up'' polarization  (which may stand
for ``cat alive'') and $d$ for the ``down'' polarization (``cat dead'').
Let us further denote the components of the initial state of the recoil
by $\Phi_k$, and its density matrix by $\beta_{k'k''}$.

The (not normalized) final-state density
matrix, $\rho_{n'l',n''l''}$, is given by the transition matrix
elements together with the initial-state density matrix:
\begin{equation}
\rho_{n'l',n''l''}~=~ \sum_{p'k'p''k''}~\langle n'l'|~T~|~p'k'\rangle~
\sigma_{p'p''}~\beta_{k'k''}~\langle~p''k''~|~T~|~n''l''~ \rangle  \quad .
\label{C11}
\end{equation}
This equation shows which interferences are possible. Thus the $u$ -- $d$
interference is possible only if the
density matrix $\sigma$ has non-vanishing off-diagonal elements.
Similarly, the possibility of any other interference is determined by
the non-vanishing of the corresponding off-diagonal elements of the
relevant {\em initial state} density matrix.

Further limitations arise from the measurement operator. The general form
of this operator interacting with the emitted particle is
\begin{equation}
{\it M}~=~\kappa_{u',u'';d',d''}~\tau_{p'p''}
~\delta(\theta - \theta')~\delta(\varphi - \varphi')
\delta(\theta - \theta'')~\delta(\varphi - \varphi'') \label{C3}
\end{equation}
The fact that the measuring operator (\ref{C3}) does not interact with
the recoiling particle leads to overlap of the wave functions $\chi_n$
in the evaluation of the measurement matrix element, which enforces
$n'~=~n''$. Evidently any possibility of $u-d$ interference hinges on the form
of the operator $\tau$: it must connect the $u$ and the $d$ states.
If instead it is of the form $\tau_{p'p''}~=~\delta_{p',p''}$ the
result of the measurement is
\begin{eqnarray}
P(\theta,\varphi)~&=&~\sum_{n'l',n''l''}~
\langle \psi_{n'l'}|~{\it M}~| \psi_{n''l''} \rangle~
\rho_{n'l',n''l''}~\delta_{n',n''}\cr
\cr
&=&~\sum_{n,l'l''}~ \phi_{l'}(\theta,\varphi)^*~ \phi_{l''}(\theta,\varphi)~
\rho_{nl',nl''}~ \quad .
\label{C5}
\end{eqnarray}
This way, not performing a measurement results in the density matrix losing the corresponding
off-diagonal elements; here $\rho_{n'l',n''l''}$ becomes $\rho_{nl',nl''}$,
i.e., diagonal in $n$. Further, interferences between the different states,
specified by the quantum numbers $l', ~l''$, arise only if the
corresponding off-diagonal elements of the final state density matrix,
Eq.~(\ref{C11}), are
non-zero, which again is possible only if $\sigma$ has non-vanishing
off-diagonal elements. The latter condition is fulfilled in electromagnetic
transitions since in the incoming photon plane wave all multipoles
are coherent. If the initial state of the system is a mixture of
non-interfering states, i.e., if the density matrix $\beta_{k'k''}$
is diagonal, the final state would consist of non-interfering blocks
of states; however, the states within each of these blocks of a given $l'$
could exhibit
full interference.

Returning to our example of the initial interaction in the measurement chain,
it is obvious that no macroscopic classical apparatus, no
matter how small, will have the simplicity of the wave function
$\psi_{nl}$, Eq.~(\ref{C2a}). The split into the classes $u$ and $d$
therefore will take place early in the chain; probably at its
first link. As discussed in the previous Section, decoherence is a very
rapid process. Even if the orthogonality of the states of the
reaction products, the equivalent to $\chi$, Eq.~(\ref{C2a}), is
incomplete, or generally, even if these states could exhibit interference,
full decoherence will arrive after a
very few links of the chain of events leading to the pointer position.
Hence the density matrix of the input to the last link before the
pointer, the analogue to the product $\sigma~\beta$ of Eq.~(\ref{C11}),
will be strictly diagonal, which is the form a classical object should have.
Consequently no interference as in Eq.~(\ref{C1}) is possible; and the
pointer behaves like a classical system. Thus, Schr\"odinger's cat will
never exhibit ``the suspended animation''; it will be either dead or
alive, independently of our state of mind.

%{\bf Note:  We may insert an Appendix here.}

\section{Summary}

A full explanation of the evolution, as well as of a measurement, of a
quantum system requires both the
particle and the wave aspects: the first to provide the ``yes-no"
decision, the second to provide the ``how much" of the
measurement.  More precisely, if the particle aspect gives the
``yes'' decision, the wave aspect provides the probability of the
particular outcome, of the particular reaction branch. Measurement or no
measurement, a fully deterministic description of the evolution is
not possible.  The attention by an experimentalist plays no role,
except that he can influence the further evolution of the system,
for example by switching off the apparatus.

The question of the interference in the results of a quantum
measurement, i.e. the question of the meaning of the superposition
of states describing different pointer positions, is answered by
investigating the meaning of a classical object in the quantum
formulation. It results in the statement that classical objects
do not exhibit interference effects. They can not be described
by wave functions; their description requires the use of density
matrices; more particularly, of diagonal or block-diagonal density
matrices. The
reasons for the conversion of the outcome of a measurement from
the quantum to the classical realm are the inevitable dissipation
associated with
the quantum measurement process, and the strength of the process
of decoherence.

Since the
particle aspects are not available within quantum mechanics  they
must be supplied ``by hand" when attempting a description within
that framework.  This artifact is called ``the collapse
of the wave function".  Quantum theory does not separate the
particle and the wave aspects.  The quantum physics field can,
however, be factorized into these aspects, cf. Eq.~(\ref{I.5}).
The wave parts of this factorization are the wave functions of
quantum mechanics; hence the correctness of the predictions of
quantum mechanics.  Still, in agreement with Einstein's observation,
quantum mechanics itself is incomplete.

%\newpage
\setcounter{equation}{0}
\renewcommand{\theequation}{A.\arabic{equation}}
\noindent
{\Large\bf Appendix A: Density Matrices}

We begin with quantum mechanics. A system which can be described by a
wave function is said to be in
a ``pure'' state. The pure state can be also defined as exhibiting full
interference. The probability of finding the particle at the point
$x$ is
\begin{equation}
P(x,t)~=~ \int~d^3x'~ \psi^*(x',t)~ \delta^3(x - x')~ \psi(x',t)   \quad ,
\label{DX1}
\end{equation}
The most general pure state is given by a wave
function which is the superposition of the eigenfunctions of the Hamiltonian:
\begin{equation}
\psi(x,t)~=~ \sum_j~ C_j~\psi_j(x)~e^{-iE_j\,t}  \label{DX2}
\end{equation}
with $\psi_j$ corresponding to energy level $E_j$.  Inserting (\ref{DX2})
in (\ref{DX1}) one sees that at every fixed
position $x$ the probability fluctuates with a superposition of
energy differences $\cos(E_j-E_{j'})t$, {\it which is not the behavior
of a classical system}, as observed in our World: a system localized
in space and time, e.g., a book on a table, is not an eigenstate of
energy or momentum, but nonetheless it does not exhibit fluctuations.
These fluctuations do not arise for a fully impure system, for which
the state is
described by a diagonal density matrix. The general matrix element of
the density-operator is
\begin{equation}
D_{kj}(x,t)~=~ \int d^3x'~\psi_j(x')~e^{-iE_j\,t}~\delta^3(x-x')~
\psi_k^*(x')~e^{iE_k\,t} \quad ; \label{DX3}
\end{equation}
for a pure state the state density matrix is
\begin{equation}
\varrho_{jk}~=~C_j~C_k^* \quad , \label{DX4}
\end{equation}
while for a fully impure (also called ``mixed'') state the state density
matrix is
\begin{equation}
\varrho_{jk}~=~\delta_{j,k}~|C_j|^2 \quad . \label{DX5}
\end{equation}
The probability is computed by tracing the product of the density operator
and the state density matrix. Thus Eq.~(\ref{DX1}) then reads
\begin{equation}
P(x,t)~=~\sum_{j,k}~\varrho_{jk}~D_{kj}(x,t)~ \label{DX6}
\end{equation}
which for a fully impure state indeed shows no interference between
the components making up the state.

{\em This way, a classical system cannot be represented by a wave function.}
It must be represented by a density matrix, which, in particular, must
be diagonal. {\em Classical systems are of necessity mixed states.}

We now define the QP state vectors appropriate to the density matrix
formalism. We shall call them ``density state vectors''.  To that end we
must augment the QM wave functions with
appropriate Fock operators. Thus the field operator is
\begin{equation}
\Psi(x,t)~=~ \sum_j~ b_j~\psi_j(x)~e^{-iE_j\,t} \quad,  \label{DX7}
\end{equation}
and the density state vector
\begin{equation}
\Omega_{jk}~=~ b^{\dagger}_j~|V \rangle~\varrho_{jk}~\langle V|~ b_k \quad ,
\label{DX9}
\end{equation}
which is bi-linear in the Fock operators. The expectation value of an
operator, $\cal O$, then is
\begin{eqnarray}
\langle \cal O \rangle &=& \sum_{j,k}~ \langle V| b_j~
\Psi^{\dagger}(x,t)~ {\cal O}(x,y)~ \Psi(y,t)~
b^{\dagger}_k |V \rangle \varrho_{jk} \cr
  &=& Tr~ \Psi^{\dagger}(x,t)~ {\cal O}(x,y)~
\Psi(y,t)~\Omega_{jk}~  \quad. \label{DX10}
\end{eqnarray}
Here the $Tr$ includes tracing over the indices $j,k$ and re-ordering
the Fock operators cyclically, without introducing commutator
phases.

\setcounter{equation}{0}
\renewcommand{\theequation}{B.\arabic{equation}}
\noindent
{\Large\bf Appendix B: More on Decoherence in Quantum Physics}

In this appendix we will illustrate the decoherence of a measured system
after interacting with a macroscopic measuring device.  Both of the
measured and the measuring are treated as quantum systems, but classically
mixed states will be realized after their interaction .  Our
decoherence result requires nothing more than the framework of
quantum physics advocated in this paper; i.e. no extra projection
or collapse of the wave function postulate.  This is possible because
quantum physics provides us the mechanism of the chain amplifying the first
interaction between the measured and the macroscopic device through
the many microscopic constituents of the device.

Explicitly, we will make use of two ingredients: (i) quantum physics
detector functions, Eqs. (21,24,34,35,53), and (ii) the many degrees
of freedom of the measuring device that are unobserved or unobservable.

To begin with we will derive a result for later use.  In the spirit
of~(6) we expand the field operator either in terms of the set of
c-number functions $\{\psi_\alpha(x,t)\}$ or another set $\{\phi_\beta(x,t)\}$,
which satisfy the same boundary condition but arbitrary otherwise (i.e.
corresponding to a different set of quantum numbers $\beta$).
\begin{eqnarray}
\Psi(x,t) &=& \sum_\alpha a_\alpha\psi_\alpha(x,t) \quad , \label{rep1}\\
&=& \sum_\beta b_\beta\phi_\beta(x,t) \quad .
\label{rep2}
\end{eqnarray}
We can take these two sets to be orthonormal without any loss of generality.
Thus,
\begin{eqnarray}
a_\alpha = \sum_\beta C_{\alpha\beta} b_\beta \quad ,\label{link}\\
\left[ a_\alpha,b_\beta^\dagger\right]_{+} = C_{\alpha\beta} \quad ,
\label{result1}
\end{eqnarray}
where
\begin{eqnarray}
C_{\alpha\beta} \equiv \int\psi^*_\alpha\phi_\beta \quad .
\end{eqnarray}

We now choose an arbitrary state $|\alpha\rangle
= a_\alpha^\dagger|V\rangle$, say, and
perform the interrogation
\begin{eqnarray}
\langle V|\Psi|\alpha \rangle &=& \psi_\alpha \quad , \nonumber\\
 &=& \sum_\beta C^*_{\alpha\beta}\phi_\beta \quad ,
\end{eqnarray}
if the expansions~(\ref{rep1}) and~(\ref{rep2}) respectively
are employed for $\Psi$.
Then from the orthonormality of $\psi$'s and $\phi$'s it follows that
$\sum_\beta \left|C_{\alpha\beta}\right|^2 = 1$, which implies
\begin{eqnarray}
\left|C_{\alpha\beta}\right| \le 1 \quad .
\label{result}
\end{eqnarray}
(The last result can also be obtained simply by considering
$\left[a_\alpha,a^\dagger_{\alpha}\right]_+ = 1 = \left[b_\beta,
b^\dagger_\beta\right]_+$ with the substitution of~(\ref{link}).)

Coming back to the decoherence problem,
let the measured system be in the pure state before the measurement,
thus its density operator is given by (for
simplicity, we restrict the number of states to 2)
\begin{eqnarray}
\rho &=& \sum_{i,j=1,2}\rho_{ij} \quad b^\dagger_i|V \rangle\langle V|b_j
\quad .
\end{eqnarray}
The density operator of the whole system is the direct product of those
of the measured and the device $\varrho$.

We now go directly to the end of the measurement process, leapfrogging
the measurement chain.  The detector function, Eq.~(\ref{M.5}), is then
replaced by the effective detector function
\begin{eqnarray}
D &=& \eta_1  B^\dagger_1b_1 + \eta_2  B^\dagger_2 b_2,
\end{eqnarray}
here $B^\dagger_j$ creates the final observable states of the aparatus,
i.e. the ``pointer position'', and
\begin{eqnarray}
\eta_{1,2} &=& \prod_{m=1}^N d^{(1,2)\dagger}_m \quad ,
\label{eta}
\end{eqnarray}
where $d^\dagger_m$ are the creation operators for the
(unobserved/unobservable) degrees of freedom in the chain, and
$m=1,\cdots,$ number of degrees of freedom $N$.
Thus the density operator after the measurement becomes
\begin{eqnarray}
\rho\otimes\varrho &\to& D\left(\rho\otimes\varrho\right)D^\dagger \quad
.
\end{eqnarray}

The reduced density operator is then obtained by tracing
over the unobserved (or unobservable) dynamical degrees of freedom of
the aparatus
\begin{eqnarray}
\rho({\rm after}) &=& {\rm Tr'}\left(D\left(\rho\otimes\varrho\right)D^\dagger
\right)\nonumber\\
&=& E_1 P_1  B^\dagger_1|V \rangle\langle V|B_1 ~+~
E_2 P_2  B^\dagger_2|V \rangle\langle V|B_2 ~+ \nonumber\\
&& \mbox{\rm interference term}.
\label{reduced}
\end{eqnarray}
In this expression, $P_i= \rho_{ii}$, $i=1,2$, is the probability that
the system is
found to be in state $i$; and
%\begin{eqnarray}
$E_i = {\rm Tr'}\left(\eta_i\varrho\eta_i^\dagger\right)$
%\end{eqnarray}
is a measure of the efficiency of the detector in detecting the state
$i$.  The interference term of~(\ref{reduced}) contains quantities that
are explicitly proportional to the trace
${\rm Tr'}\left(\eta_i\varrho\eta_j^\dagger\right)$,
where $i\not=j$.  And with the representation~(\ref{eta}) for $\eta_i$,
a typical term of this trace has the form
\begin{eqnarray}
{\rm Tr'}\left(\eta_i\varrho\eta_j^\dagger\right) &\sim&
\prod_{m=1}^N\prod_{n=1}^N ~\langle |d^{(j)}_m d^{(i)\dagger}_n|\rangle
~+ \quad \mbox{\rm similar terms} \quad ,
\label{interfere}
\end{eqnarray}
where $|\rangle = d^\dagger \cdots d^\dagger|V\rangle$ represent the
initial states
of the many constituents of the detector.

(Anti-)commuting the $d$-annihilation operators in~(\ref{interfere}) to
the right and using the results in~(\ref{result1},\ref{result}), which now
translate to a strict inequality because $i\not= j$,
\begin{eqnarray}
\left|\left[d^{(j)}_m,d^{(i)\dagger}_n\right]_+\right| &<& 1 \quad ,
\end{eqnarray}
we finally obtain the result that the off-diagonal elements of the reduced
density operator vanish
\begin{eqnarray}
{\rm Tr'}\left(\eta_i\varrho\eta_j^\dagger\right)
&\stackrel{N\to\infty}{\longrightarrow}& 0 \quad , \quad i\not= j \quad
,
\end{eqnarray}
as the terms on the right hand side of~(\ref{interfere}) are products of
complex numbers of magnitudes less than one.
That is, with a macroscopic measuring device, we have complete decoherence
\begin{eqnarray}
\rho({\rm after})
&\stackrel{N\to\infty}{\longrightarrow}&
E_1 P_1 B^\dagger_1|V \rangle\langle V|B_1 ~+~
E_2 P_2 B^\dagger_2|V \rangle\langle V|B_2 \quad .
\end{eqnarray}
The estimation for the time required for complete decoherence is more
complicated but may be calculated with the full use of quantum field theory
or its non-relativistic equivalent.

%\newpage
%\setcounter{equation}{0}
%\renewcommand{\theequation}{C.\arabic{equation}}
\noindent
{\Large\bf Appendix C: Preparation of State}

A conceptual and semantic confusion exists between the similar but
distinct processes of ``measurement'' and ``preparation of a system
in a specific state''.  Namely, ``the preparation of the initial
state for an experiment'' consists in allowing the desired state to
remain and rejecting all other states.  This supposedly is achieved
by an appropriate measurement.  The selection of the desired state
then in quantum mechanics is described  as the ``collapse of the
wave function'':
the ``measurement puts the state into an eigenstate of
the measuring apparatus''.  The prototype of such a setup is a
Stern-Gerlach apparatus.  Indeed, the atoms emerge  from the
source in a beam in which they are a mixture of all possible
states, of which, say, there are N.  At the output of the
Stern-Gerlach
apparatus then emerge N separated beams, one each for the
different states.  Now one simply must supply a collimator such
that only the beam containing the desired state is transmitted; all
other beams are absorbed.  No interaction with the transmitted
particles has taken place; this process is non-dissipative. The
statement that ``the measurement puts the system into an eigenstate
of the apparatus'' in fact is not accurate since actually no
measurement has been performed on
the particle.  We only know that if a particle emerges in that
output beam, then it has that particular polarization -- except
for background effects arising from collimator scattering etc.

Another possibility for preparing the system in a specified state
is the technique called ``tagging'', which is of the kind of the
EPR setup~\cite{4}: measurement of the characteristics of one
partner in a correlated pair of particles provides information on
the state of
the other partner.  A \hbox{well-known} example is that of the
tagged bremsstrahlung photons: if the detector, which is an
electron detector, registers, i.e., measures the characteristics
of, a recoiling electron then one ``knows'' that the  other
particle, here the bremsstrahlung photon, is in a well-defined
state having definite energy, direction of propagation, and
polarization.  Again, no measurement has been performed on the
tagged photon, which is the particle of interest.

Overall, since in the preparation of the state the system of
interest has actually been left alone, this process should not be
denoted as ``a measurement.''  A more precise term would be that the
preparation is ``a filtering action.''  In quantum physics the
filtering is accomplished either directly by absorbing the
particles being in the undesired states  or by performing a
measurement on another part of the system, in an \hbox{EPR-type}
arrangement.  Any measurement directly on the system of interest
would be of the kind described above in the context of the
\hbox{two-slit} experiment when checking for the passage of the
particle through the slit, and involves an interaction of the
system with an apparatus. The re-emitted system then inescapably is
in a new state, and has new not necessarily known properties.

%\newpage

\end{document}